\batchmode
\documentclass[12pt]{article}    
\usepackage{hyperref}
\usepackage{graphics}
\usepackage{epsfig}
\usepackage{natbib}   
\usepackage{graphicx} 
\usepackage{amssymb}  
\usepackage{multirow} 
\usepackage{epstopdf}
\usepackage{bm}
\usepackage{amsmath}
\usepackage{amsfonts}
\usepackage{amssymb}
\usepackage{amssymb, longtable}
\usepackage{pdflscape}

\begin{document}
\newtheorem{definition}{Definition}
\newtheorem{theorem}{Theorem}
\newtheorem{example}{Example}
\newtheorem{corollary}{Corollary}
\newtheorem{lemma}{Lemma}
\newtheorem{proposition}{Proposition}
\newenvironment{proof}{{\bf Proof:\ \ }}{\qed}
\newcommand{\qed}{\rule{0.5em}{1.5ex}}
\newcommand{\bfg}[1]{\mbox{\boldmath $#1$\unboldmath}}

\begin{center}

\section*{Risks aggregation in multivariate dependent Pareto distributions}

\vskip 0.2in {\sc \bf Jos\'e Mar\'{\i}a Sarabia$^a$\footnote{E-mail addresses: sarabiaj@unican.es (J.M. Sarabia).}, Emilio G\'omez-D\'eniz$^b$ \\ Faustino Prieto$^a$, Vanesa Jord\'a$^a$}

\vskip 0.1in

$^a$Department of Economics, University of Cantabria, Avda de los Castros s/n, 39005-Santander, Spain

\vskip 0.1in

$^b$Department of Quantitative Methods in Economics and TiDES Institute, University of Las Palmas de Gran Canaria, 35017-Las Palmas de G.C., Spain
\end{center}

\begin{abstract}\noindent
In this paper we obtain closed expressions for the probability distribution function, when we consider aggregated risks with multivariate dependent Pareto distributions. We work with the dependent multivariate Pareto type II proposed by Arnold (1983, 2015), which is widely used in insurance and risk analysis. We begin with the individual risk model, where we obtain the probability density function (PDF), which corresponds to a second kind beta distribution. We obtain several risk measures including the VaR, TVaR and other tail measures. Then, we consider collective risk model based on dependence, where several general properties are studied. We study in detail some relevant collective models with Poisson, negative binomial and logarithmic distributions as primary distributions. In the collective Pareto-Poisson model, the PDF is a function of the Kummer confluent hypergeometric function, and in the Pareto-negative binomial is a function of the Gauss hypergeometric function. Using the data set based on one-year vehicle insurance policies taken out in 2004-2005 (Jong and Heller, 2008), we conclude that our collective dependent models outperform the classical collective models Poisson-exponential and geometric-exponential in terms of the AIC and CAIC statistics.
\end{abstract}

\noindent \textit{Key Words}: Dependent risks; individual risk model; collective risk model.

\section{Introduction}

In the context of the classical theory of risk, consider the individual and the collective risk models (see Kass et al. (2001) and Klugman et al. (2008)). These models rely on the assumptions of the independence between: (i) different claim amounts; (ii) the number of claims and claim amounts and (iii) claim amounts and interclaim times.

The use of the independence hypothesis facilitates the computation of many risks measures, but can be restrictive in different contexts. The need for generalizations in both individual and collective classical models (by considering some kind of dependence structure) has led several researches to model dependence.

In this context, Sarabia and Guill\'en (2008) have considered extensions of the classical collective model assuming that the conditional distributions $S|N$ and $N|S$ belong to some prescribed parametric family, where $S$ is the total claim amount and $N$ is the number of claims. G\'omez-D\'eniz and Calder\'in-Ojeda (2014), using conditional specification techniques, have obtained discrete distributions to be used in the collective risk model to compute the right-tail probability of the aggregate claims size distribution.

Albrecher and Teugels (2006) have considered a dependence structure based on a copula for the interclaim time and the subsequent claim size. Boudreault et al. (2006) studied an extension of the classical compound Poisson risk model, where the distribution of the next claim amount was defined as a function of the time elapsed since the last claim. Cossette et al. (2008) have considered another extension introducing a dependence structure between the claim amounts and the interclaim time using a generalized Farlie-Gumbel-Morgenstern copula. Cossette et al. (2004) have considered a variation of the compound binomial model in a markovian environment, which is an extension of the model presented by Gerber (1998). Compound Poisson approximations for individual models with dependent risk was considered by Genest et al (2003).

A recent model of dependence between risks has been proposed by Cossette et al. (2013). These authors consider a portfolio of dependent risks whose multivariate distribution is defined with the Farlie-Gumbel-Morgenstern copula, with mixed Erlang distribution marginals.

In this paper we obtain closed expressions for the probability distribution function modeling aggregated risks with multivariate dependent Pareto distributions between the different claim amounts. We work with the dependent multivariate Pareto type II proposed by Arnold (1983, 2015), which is widely used in insurance and risk analysis. We begin with the classical individual risk model. For this model we obtain the probability density function (PDF), which corresponds to a second kind beta distribution. We also obtain several risk measures including the VaR, TVaR and other tail measures. Then, we consider collective risk model based on dependence, whose main general properties are studied. We study in detail some relevant collective models with Poisson, negative binomial and logarithmic distributions as primary distributions. For these three models, we obtain simple and closed expressions for the aggregated distributions.

The contents of this paper are the following. In Section \ref{section2} we present the main univariate distributions used in the paper. The class of multivariate dependent Pareto distribution considered to model aggregated risks is presented in \ref{section3}. Section \ref{section4} presents the individual risk model under dependence and Section \ref{section5} introduces the collective risk model under dependence. After presenting general results we study the compound models where the primary distribution is Poisson, negative binomial, geometric and logarithmic and the secondary distribution is Pareto. Section \ref{section6} includes an example with real data. The conclusions of the paper are given in Section \ref{section7}.

\section{Univariate distributions}\label{section2}

In this section, we introduce several univariate random variables which will be used in the paper.

We work with the Pareto distribution with PDF given by,
\begin{equation}\label{PDFPareto}
f(x;\alpha,\beta)=\frac{\alpha}{\beta(1+x/\beta)^{\alpha+1}},\;\;x>0,
\end{equation}
and $f(x;\alpha,\beta)=0$ if $x<0$, where $\alpha,\beta>0$. Here, $\alpha$ is a shape parameter and $\beta$ is a scale parameter. We represent $X\sim {\cal P}a(\alpha,\beta)$.

We denote by $X\sim {\cal G}a(\alpha)$ a gamma random variable with PDF $f(x)=\frac{x^{\alpha-1}e^{-x}}{\Gamma(\alpha)}$ if $x>0$, whith $\alpha>0$. The exponential distribution with mathematical expectation 1 is denoted by ${\cal G}a(1)$.

The following lemma provides a simple stochastic representation of the Pareto distribution as quotient of random variables. The proof is straightforward and will be omitted.
\begin{lemma}\label{lemma1}
Let $U_1$ and $U_\alpha$ independent gamma random variables such that $U_1\sim {\cal G}a(1)$ and $U_\alpha\sim {\cal G}a(\alpha)$, where $\alpha>0$. If $\beta>0$, the random variable,
\begin{equation}
X=\beta\frac{U_1}{U_\alpha}\sim {\cal P}a(\alpha,\beta).
\end{equation}
\end{lemma}

An extension of the Pareto distribution (\ref{PDFPareto}) is the following. A random variable $X$ is said to be a second kind beta distribution if its PDF is of the form,
\begin{equation}\label{PDFsecondkindbeta}
f(x;p,q,\beta)=\frac{x^{p-1}}{\beta^pB(p,q)(1+x/\beta)^{p+q}},\;\;x>0,
\end{equation}
and $f(x;p,q,\beta)=0$ if $x<0$, where $p,q,\beta>0$ and $B(p,q)=\frac{\Gamma(p)\Gamma(q)}{\Gamma(p+q)}$ denotes the beta function. This random variable corresponds to the Pearson VI distribution in the classical Pearson systems of distributions and we represent $X\sim {\cal B}2(p,q,\lambda)$. If we set $p=1$ in (\ref{PDFsecondkindbeta}), we obtain a Pareto distribution ${\cal P}a(q,\beta)$ like (\ref{PDFPareto}).

The second kind beta distribution has a simple stochastic representation as a ratio of gamma random variables. If $U_p$ and $U_q$ are independent gamma random variables, the new random variable $X=\beta\frac{U_p}{U_q}$ has the PDF defined in (\ref{PDFsecondkindbeta}).

\section{The multivariate Pareto class}\label{section3}

Now we present the class of multivariate dependent Pareto distribution which will be used in the different models.

In the literature several classes of multivariate Pareto distributions have been proposed. One of the main classes was introduced by Arnold (1983, 2015), in the context of the hierarchy Pareto distributions proposed by this author. Other classes were proposed by Chiragiev and Landsman (2009) and Asimit et al. (2010). The conditional dependence structure is the base of the construction of the proposals by Arnold (1987) and Arnold et al. (1993) (see also Arnold et al, 2001), where two different dependent classes are obtained.

\begin{definition}\label{definitio1}
Let $Y_1,Y_2,\dots,Y_n$ and $Y_\alpha$ be mutually independent gamma random variables with distributions $Y_i\sim {\cal G}a(1)$, $i=1,2,\dots,n$ and $Y_\alpha\sim {\cal G}a(\alpha)$ with $\alpha>0$. The multivariate dependent Pareto distribution is defined by the stochastic representation,
\begin{equation}\label{multiPareto}
\bm X=\left(X_1,X_2,\dots,X_n\right)^\top=\left(\beta\frac{Y_1}{Y_\alpha},\beta\frac{Y_2}{Y_\alpha},\dots,\beta\frac{Y_n}{Y_\alpha}\right)^\top,
\end{equation}
where $\beta>0$.
\end{definition}

Note that the common random variable $Y_\alpha$ introduces the dependence in the model.

\subsection{Properties of the multivariate Pareto class}

We describe several properties of the multivariate Pareto defined in (\ref{definitio1}).

\begin{itemize}
\item Marginal distributions. By construction, the marginal distributions are Pareto,
$$
X_i\sim {\cal P}a(\alpha,\beta),\;\;i=1,2,\dots,n.
$$
\item The joint PDF of the vector $\bm X$ is given by,
\begin{equation}\label{jointPDFPareto}
f(x_1,\dots,x_n;\alpha,n)=\frac{\Gamma(\alpha+n)}{\Gamma(\alpha)\beta^n}\frac{1}{(1+\sum_{i=1}^nx_i/\beta)^{\alpha+n}},\;\;x_1,\dots,x_n>0.
\end{equation}
This expression corresponds to the multivariate Pareto type II proposed by Arnold (1983, 2015).
\item The covariance matrix is given by,
$$
cov(X_i,X_j)=\frac{\beta^2}{(\alpha-1)^2(\alpha-2)},\;\;\alpha>2,\;\;i\neq j
$$
and the correlation between components is,
$$
\rho(X_i,X_j)=\frac{1}{\alpha},\;\;\alpha>2,\;\;i\neq j
$$
\item General moments. The moments of (\ref{definitio1}) are,
$$
E[X_1^{r_1}\cdots X_n^{r_n}]=\frac{\Gamma(\alpha-A)}{\Gamma(\alpha)}\prod_{i=1}^n\beta^{r_i}\Gamma(1+r_i),
$$
where $A=r_1+\dots+r_m$ and $\alpha>A$.
\end{itemize}

The dependence structure of $\bm X$ is studied in the following result
\begin{proposition}\label{proposition1}
The random variables $\bm X=(X_1,\dots,X_n)^\top$ are associated, and then $\mbox{cov}(X_i,X_j)\ge 0$, if $i\neq j$.
\end{proposition}
\begin{proof}
See Appendix.
\end{proof}\\

\noindent{\bf Remark:} Let us consider the multivariate Pareto survival function of (\ref{jointPDFPareto}) given by,
$$\bar F(x_1,\dots,x_n)=\left(1+\sum_{i=1}^n\frac{x_i}{\beta}\right)^{-\alpha},\;\;x_1,\dots,x_n>0,$$
with $\alpha,\beta>0$.
For this family, the associated copula is the Pareto copula or Clayton copula,
$$
C(u_1,\dots,u_n;\alpha)=\left(u_1^{-1/\alpha}+\dots+u_n^{-1/\alpha}-n+1\right)^{-\alpha}.
$$
Note that the dependence increases with $\alpha$, being the independence case  obtained when $\alpha\to 0$ and the Fr\'echet upper bound when $\alpha\to\infty$ .

\section{The individual risk model under Pareto dependence}\label{section4}

In this section we consider the individual risk model assuming dependence between risks. The distribution of sums of iid Pareto distributions was obtained by Ramsay (2006). Let $(X_1,\dots,X_n)^\top$ be the multivariate Pareto distribution defined in (\ref{multiPareto}). Then, we consider the aggregate risks $S_n=X_1+\dots+X_n$. We have the following result.
\begin{theorem}\label{theorem1}
The PDF of the aggregate random variable $S_n$, where the components are Pareto defined in (\ref{definitio1}) is given by,
\begin{equation*}\label{PDFindividual}
f_{S_n}(x;n,\alpha,\beta)=\frac{x^{n-1}}{\beta^nB(n,\alpha)(1+x/\beta)^{n+\alpha}},\;\;x>0
\end{equation*}
end $f_{S_n}(x;n,\alpha,\beta)=0$ if $x<0$, that is $S_n\sim {\cal B}2(n,\alpha,\beta)$.
\end{theorem}
\begin{proof}
See Appendix.
\end{proof}

\begin{figure}[h]
\begin{center}
\includegraphics*[scale=0.70]{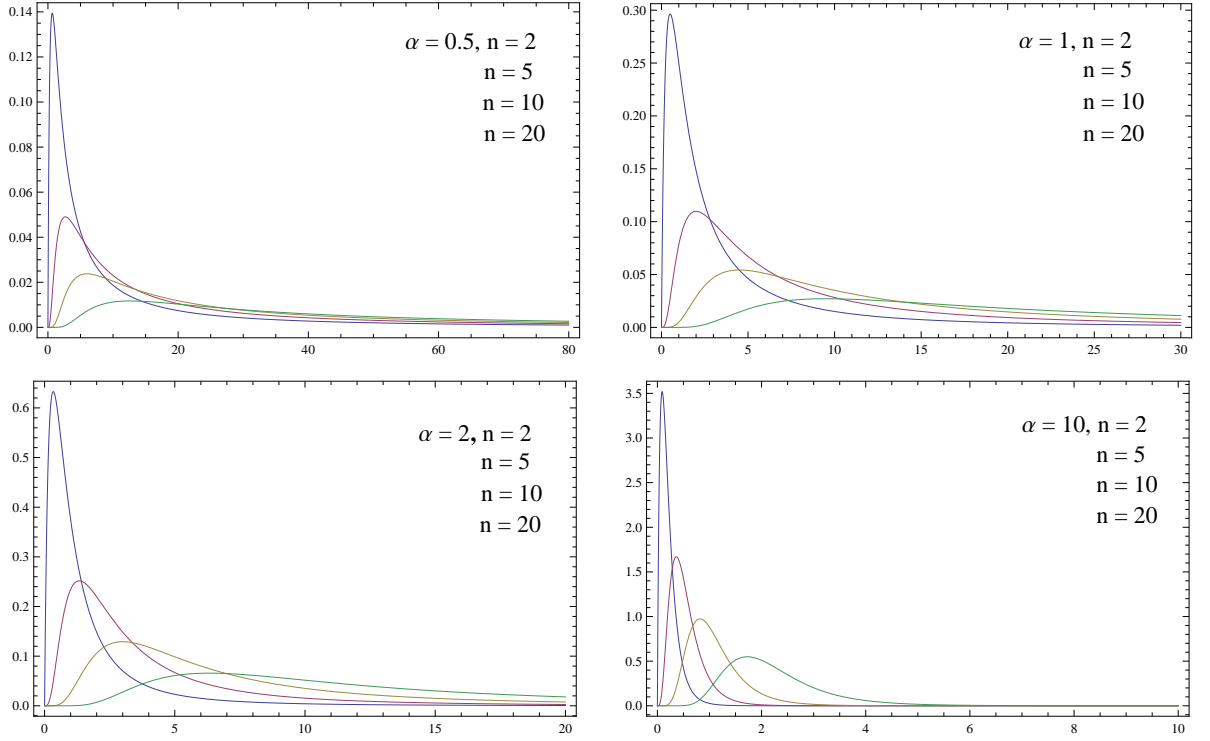}
\caption{\label{Figure1} PDF of the individual model $S_n$ for some selected values of $\alpha$ and $n$.}
\end{center}
\end{figure}

Note that (\ref{PDFindividual}) corresponds to the PDF of a second kind beta distribution defined in (\ref{PDFsecondkindbeta}). Figure \ref{Figure1} represents the PDF (\ref{PDFindividual}) for $\alpha=1/2; 1; 2$ and $10$ for $n=2,5,10$ and $20$.

\subsection{Risk measures}

Here we present some risk measures for the second kind beta distribution, which can be applied for the aggregate PDF given in (\ref{PDFindividual}). The value at risk VaR at level $u$, with $0<u<1$ of a random variable $X$ with CDF $F(x)$ is defined as,
$$\mbox{VaR}[X;u]=\inf\{x\in \mathbb{R},\,F(x)\ge u\}.$$

If $X\sim {\cal B}2(p,q,\beta)$, Guill\'en et al. (2013) have obtained that,
\begin{equation}\label{varbeta2}
\mbox{VaR}[X;u]=\beta\frac{IB^{-1}(u;p,q)}{1-IB^{-1}(u;p,q)},
\end{equation}
where $IB^{-1}(u;p,q)$ denotes the inverse of the incomplete ratio beta function, which corresponds to the quantile function of the classical beta distribution of  the first kind. Then, using (\ref{varbeta2}) for the aggregate distribution (\ref{PDFindividual}) we have
$$\mbox{VaR}[S_n;u]=\beta\frac{IB^{-1}(u;n,\alpha)}{1-IB^{-1}(u;n,\alpha)},$$
with $0<u<1$.

The following result provides higher moments of the worst $x_u$ events, which extend popular risk measures.
\begin{lemma}\label{lemma3}
Let $X\sim {\cal B}2(p,q,\beta)$ be a second kind beta distribution. Then, the conditional tail moments are given by,
\begin{equation}\label{highermoments}
E[X^r|X>x_u]=\frac{\beta^r\Gamma(p+r)\Gamma(q-r)}{(1-u)\Gamma(p)\Gamma(q)}\left\{1-B\left(\frac{x_u/\beta}{1+x_u/\beta};p+r,q-r\right)\right\},
\end{equation}
where $x_u=VaR[X;u]$ represents the value at risk with $u\in (0,1)$.
\end{lemma}
\begin{proof}
See Appendix.
\end{proof}

If we take $r=1$ in (\ref{highermoments}) we obtain the tail value at risk TVaR, which was obtained by Guill\'en et al. (2013).

\section{The collective risk model under dependence}\label{section5}

In this section we consider the collective model based on dependence between claim amounts. Let $N$ be the number of claims in a portfolio of policies in a time period. Let $X_i$, $i=1,2,\dots$ be the amount of the $i$th claim and $S_N=X_1+\dots+X_N$ the aggregate claims of the portfolio in the time period considered.

\subsection{General properties}

We consider two assumptions: (1) We assume that all the claims $X_1,X_2,\dots,X_n,\dots$ are dependent random variables with the same distribution and (2) The random variable $N$ is independent of all claims $X_1,X_2,\dots,X_n,\dots$

\begin{theorem}\label{theorem2}
Let $(X_1,X_2,\dots,X_n,\dots)$ be dependent random variables with common CDF $F(x)$, and let $N$ be the observed number of claims, with PMF given by $p_n=\Pr(N=n)$, for $n=0,1,\dots$, which is independent of all $X_i$'s, $i=1,2,\dots$ Then, the CDF of the aggregate losses $S_N$ is,
\begin{equation*}
F_{S_N}(x)=\sum_{n=0}^\infty p_nF_X^{(n)}(x),
\end{equation*}
where $F_X^{(n)}(x)$ represent the CDF of the convolution of the $n$ dependent claims $(X_1,\dots,X_n)$.
\end{theorem}
\begin{proof}
See Appendix.
\end{proof}

The mean and the variance of the collective model can be found in the following result.
\begin{lemma}\label{lemma2}
The mean and the variance of $S_N$ under dependence are given by,
{\small
\begin{eqnarray}
E(S_N)&=&E(N)E(X),\label{meanSN}\\
var(S_N)&=&E(N)var(X)+var(N)(E(X))^2+E[N(N-1)]cov(X_i,X_j)\label{varSN}
\end{eqnarray}}
\end{lemma}
\begin{proof}
See Appendix.
\end{proof}

If the claims $X_i$ and $X_j$ are independent $cov(X_i,X_j)=0$ and then (\ref{varSN}) becomes in the usual formula for $var(S_N)$.

On the other hand, if the random variables $\{X_i\}$ are associated,
$$\mbox{var}(S_N^{(I)})<\mbox{var}(S_N^{(D)}),$$
where $\mbox{var}(S_n^{(I)})$ is the variance in the independent case and $\mbox{var}(S_n^{(D)})$ the variance in the dependent case. This fact is a consequence of the associated property, which leads to a positive variance between $X_i$ and $X_j$.

\subsection{Compound Pareto models}

In this section we obtain the distribution of the compound collective model $S_N=X_1+X_2+\dots+X_N$ where the secondary distribution is Pareto given by (\ref{multiPareto}), and several primary distribution are considered for $N$.

In the case of independence between claims $X_i$ this distribution was studied by Ramsay (2009) when the primary distribution is Poisson and negative binomial cases.

\subsubsection{The compound Pareto-Poisson distribution}

We consider the model where the primary distribution is a Poisson distribution.
\begin{theorem}\label{theorem3}
If we assume a Poisson distribution with parameter $\lambda$ as primary distribution and $(X_1,\dots,X_n)$ is defined in (\ref{definitio1}), the PDF of the random variable $S_N$ is given by,
\begin{equation}\label{PDFParetoPoisson}
f_{S_N}(x;\alpha,\lambda,\beta)=\frac{\alpha\lambda e^{-\lambda}}{\beta(1+x/\beta)^{\alpha+1}}{}_1F_1\left[1+\alpha;2;\frac{\lambda x/\beta}{1+x/\beta}\right],\;\;x>0
\end{equation}
and $f_{S_N}(0;\alpha,\lambda,\beta)=e^{-\lambda}$
where ${}_1F_1(a;b;z)$ denotes the Kummer confluent hypergeometric function defined by,
\begin{equation*}
{}_1F_1[a;b;z]=\sum_{n=0}^\infty\frac{(a)_nz^n}{(b)_nn!},
\end{equation*}
where $(a)_n$ represent the Pochhammer symbol defined by $(a)_n=a(a-1)\dots (a-n+1)$.
\end{theorem}
\begin{proof}
See Appendix.
\end{proof}

Figure \ref{Figure2} represents the PDF (\ref{PDFParetoPoisson}) Pareto-Poisson for some values of $\alpha=0,5; 1; 2$ and 10 and $\lambda=2,5,10$ and 20, taking $\beta=1$.

\begin{figure}[h]
\begin{center}
\includegraphics*[scale=0.70]{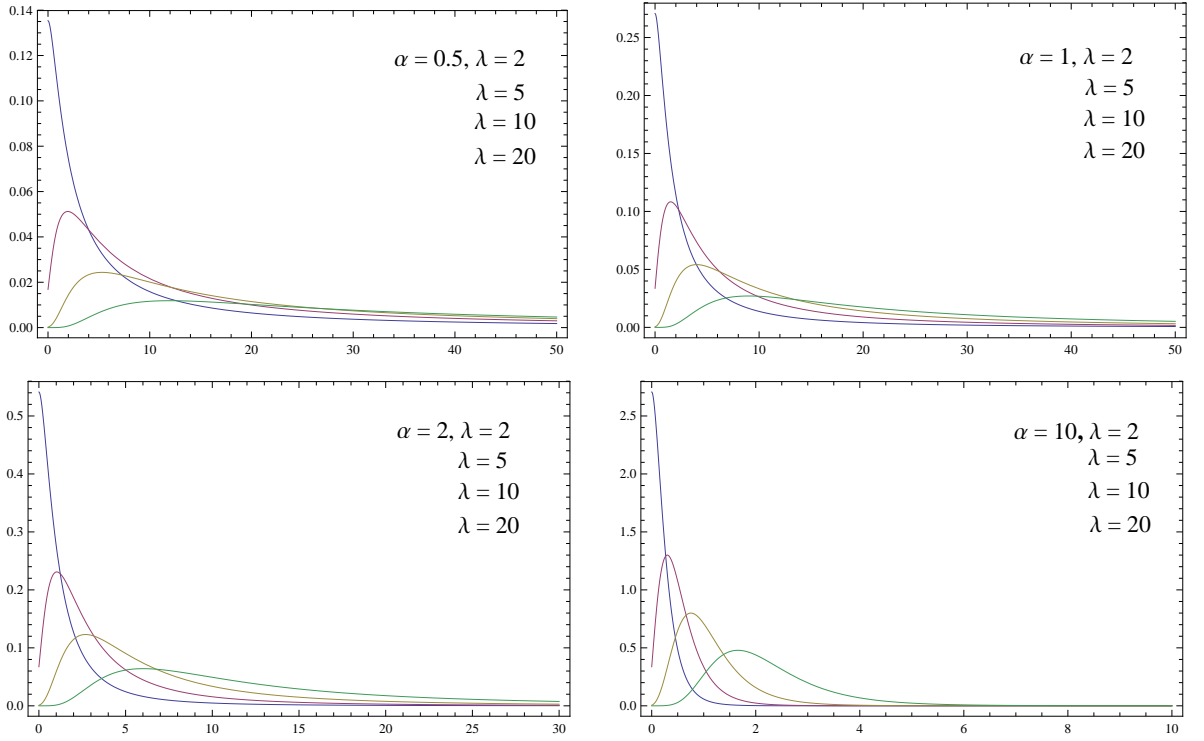}
\caption{\label{Figure2} PDF of the collective model Pareto-Poisson for $\beta=1$ and selected values of $\alpha$ and $\lambda$.}
\end{center}
\end{figure}

Using formulas (\ref{meanSN}) and (\ref{varSN}), the mean and variance of the dependent Pareto-Poisson collective model are given by,
\begin{eqnarray*}
E(S_N)&=&\frac{\lambda\beta}{\alpha-1},\;\;\alpha>1,\\
var(S_N)&=&\frac{\alpha\beta^2(\lambda+2\alpha-2)}{(\alpha-1)^2(\alpha-2)},\;\;\alpha>2.
\end{eqnarray*}

\subsubsection{The compound Pareto-negative binomial distribution}

Let $N$ be a negative binomial distribution with PMF,
\begin{equation}\label{binneg}
\Pr(N=n)=\frac{\Gamma(n+r)}{\Gamma(n+1)\Gamma(r)}p^r(1-p)^n,\;\;n=0,1,2,\dots
\end{equation}
We have the following theorem.

\begin{theorem}\label{theorempbn}
If we assume a negative binomial distribution with parameter $r$ and $p$ and PMF given by (\ref{binneg}) as primary distribution and $(X_1,\dots,X_n)$ is defined in (\ref{multiPareto}), the PDF of the random variable $S_N$ is given by,
\begin{equation*}
f_{S_N}(x;\alpha,\beta,r,p)=\frac{r\alpha(1-p)p^r}{\beta(1+x/\beta)^{\alpha+1}}{}_2F_1\left[1+r,1+\alpha;2;\frac{(1-p) x/\beta}{1+ x/\beta}\right],\;\;x>0
\end{equation*}
and $f_{S_N}(0;\alpha,\beta,r,p)=p^r$, where ${}_2F_1(a,b;c;z)$ denotes the Gauss hypergeometric function defined by,
\begin{equation*}
{}_2F_1[a,b;c;z]=\sum_{n=0}^\infty\frac{(a)_n(b)_nz^n}{(c)_nn!},
\end{equation*}
where $(a)_n$ denotes the Pochhammer symbol.
\end{theorem}

The proof is similar to the proof in Theorem \ref{theorem3} and will be omitted.

Again, using formulas (\ref{meanSN}) and (\ref{varSN}), the mean and variance of the dependent Pareto-negative binomial collective model are given by,
\begin{eqnarray*}
E(S_N)&=&\frac{r(1-p)\beta}{p(\alpha-1)},\;\;\alpha>1,\\
var(S_N)&=&\frac{r(1-p)\beta^2((1+p)(\alpha-1)+r(1-p))}{p^2(\alpha-1)^2(\alpha-2)},\;\;\alpha>2.
\end{eqnarray*}

\subsubsection{The compound Pareto-geometric distribution}

In this result we obtain the PDF of the Pareto-geometric distribution.

\begin{corollary}\label{corolario11}
If we assume a geometric distribution with parameter and PMF given by $\Pr(N=n)=p(1-p)^n$, $n=0,1,\dots$ as primary distribution and $(X_1,\dots,X_n)$ is defined in (\ref{multiPareto}), the PDF of the random variable $S_N$ is given by,
\begin{equation}\label{PDFParetogeometric}
f_{S_N}(x;\alpha,\beta,p)=\frac{\alpha p(1-p)}{\beta(1+px/\beta)^{\alpha+1}},\;\;x>0
\end{equation}
and $f_{S_N}(0;\alpha,\beta,p)=p$.
\end{corollary}

\begin{figure}[h]
\begin{center}
\includegraphics*[scale=0.70]{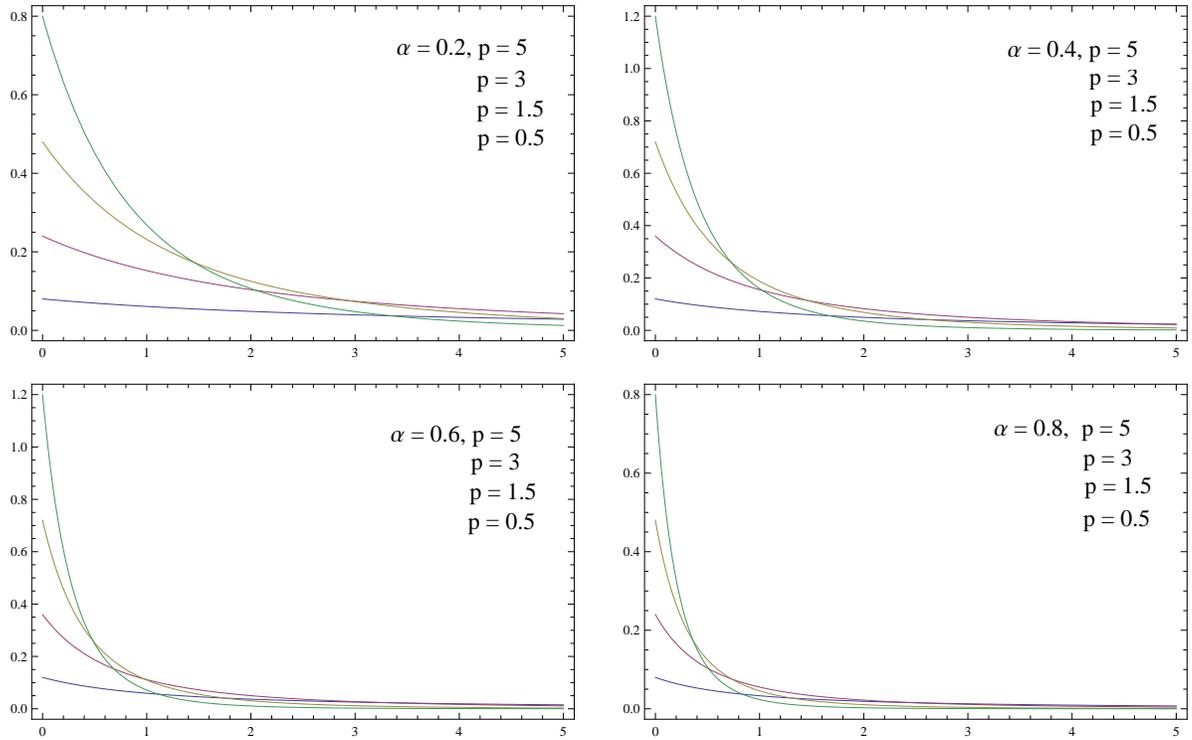}
\caption{\label{Figure3} PDF of the collective model Pareto-Geometric for $\beta=1$ and selected values of $\alpha$ and $p$.}
\end{center}
\end{figure}

Figure \ref{Figure3} represents the PDF (\ref{PDFParetogeometric}) for some selected values of $\alpha$ and $p$ ($\beta=1$).

\subsubsection{The compound Pareto-logarithmic distribution}

Let $N$ be a discrete logarithmic distribution with PMF,
\begin{equation}\label{logarit}
\Pr(N=n)=-\frac{1}{\log(1-\theta)}\frac{\theta^n}{n},\;\;n=1,2,\dots,
\end{equation}
where $\theta\in (0,1)$.

We have the following theorem.
\begin{theorem}
If we assume a logarithmic distribution with parameter $\theta$ and PMF given by (\ref{logarit}) as primary distribution and $(X_1,\dots,X_n)$ is defined in (\ref{definitio1}), the PDF of the random variable $S_N$ is given by,
\begin{equation}\label{PDFParetologarit}
f_{S_N}(x;\alpha,\beta,\theta)=-\frac{1}{\log(1-\theta)}\left[\frac{1}{x(1+(1-\theta)x/\beta)^\alpha}-\frac{1}{x(1+x/\beta)^\alpha}\right],\;\;x>0
\end{equation}
and $f_{S_N}(x;\alpha,\beta,\theta)=0$, if $x<0$.
\end{theorem}

The proof of this result is omitted.

\begin{figure}[h]
\begin{center}
\includegraphics*[scale=0.70]{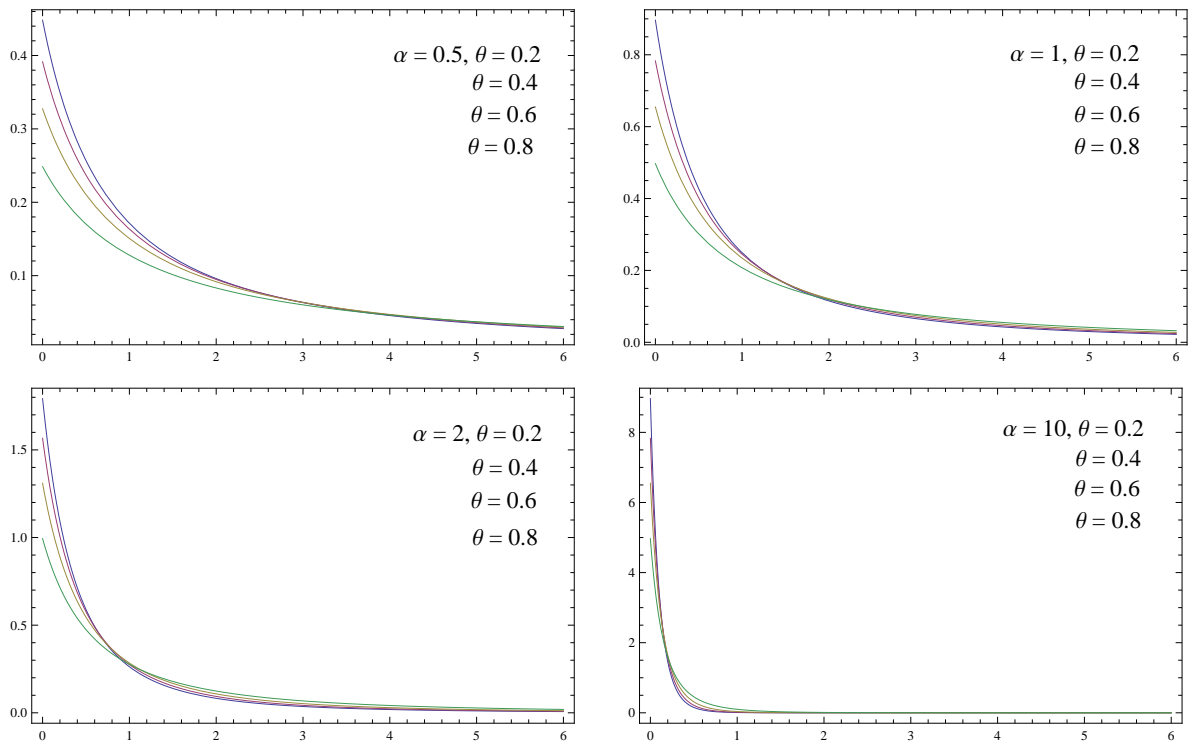}
\caption{\label{Figure5} PDF of the collective model Pareto-Logarithmic for $\beta=1$ and selected values of $\alpha$ and $\theta$.}
\end{center}
\end{figure}

Figure \ref{Figure5} represents the PDF (\ref{PDFParetologarit}) Pareto-Logarithmic for some values of $\alpha=0,5; 1; 2$ and 10 and $\theta=0.2; 0.4; 0.6$ and $0.8$ taking $\beta=1$.

The mean and variance of the dependent Pareto-Logarithmic collective model are
\begin{eqnarray*}
E(S_N)&=&\frac{a\beta\theta}{(\alpha-1)(1-\theta)},\;\;\alpha>1,\\
var(S_N)&=&\frac{a\beta^2\theta(2(\alpha-1)-\alpha\theta(1+a)+\theta(1+2a)}{(\alpha-1)^2(\alpha-2)(1-\theta)^2},\;\;\alpha>2,
\end{eqnarray*}
being $a=-1/\log(1-\theta)$.

\section{A Numerical application with real data}\label{section6}

In order to compare the performance of the models presented in this
 paper, we examined a data set based on one-year vehicle insurance
policies taken out in 2004 or 2005. This data set is available on the
website of the Faculty of Business and Economics, Macquarie University
(Sydney, Australia) (see also Jong and Heller (2008)). The first 100
observations of this data set are shown in Table \ref{tab1}, with
the following elements: from left to right, the policy number, the
number of claims and the size of the claims. The total portfolio
contains 67856 policies of which 4624 have at least one claim.
Some descriptive statistics for this data set are shown in Table
\ref{tab2}. It can be seen that the standard deviation is very
large for the size of the claims, which means that a premium based
only on the mean size of the claims is not adequate for computing
the bonus-malus premiums. The covariance between the claims and
sizes is positive and takes the value 141.574.
\renewcommand{\baselinestretch}{1}
\begin{table}[htb]\footnotesize\caption{First 100 observations of the data set\label{tab1}}
\begin{center}
\begin{tabular}{|rrr|rrr|rrr|rrr|rrr|}\hline\hline
 1 & 0 & 0     &    21 & 0 & 0 &  41 & 2 & 1811.71 & 61 & 0 & 0               & 81 & 0 & 0 \\
 2 & 0 & 0     &    22 & 0 & 0 &  42 & 0 & 0       & 62 & 0 & 0               & 82 & 0 & 0 \\
 3 & 0 & 0      &   23 & 0 & 0 &  43 & 0 & 0       & 63 & 0 & 0               & 83 & 0 & 0 \\
 4 & 0 & 0      &   24 & 0 & 0 &  44 & 0 & 0       & 64 & 0 & 0               & 84 & 0 & 0 \\
 5 & 0 & 0       &  25 & 0 & 0 &  45 & 0 & 0       & 65 & 1 & 5434.44         & 85 & 0 & 0 \\
 6 & 0 & 0       &  26 & 0 & 0 &  46 & 0 & 0       & 66 & 1 & 865.79          & 86 & 0 & 0 \\
 7 & 0 & 0       &  27 & 0 & 0 &  47 & 0 & 0       & 67 & 0 & 0               & 87 & 0 & 0 \\
 8 & 0 & 0      &   28 & 0 & 0 &  48 & 0 & 0       & 68 & 0 & 0               & 88 & 0 & 0 \\
 9 & 0 & 0      &   29 & 0 & 0 &  49 & 0 & 0       & 69 & 0 & 0               & 89 & 0 & 0 \\
 10 & 0 & 0     &   30 & 0 & 0 &  50 & 0 & 0       & 70 & 0 & 0               & 90 & 0 & 0 \\
 11 & 0 & 0     &   31 & 0 & 0 &  51 & 0 & 0       & 71 & 0 & 0               & 91 & 0 & 0 \\
 12 & 0 & 0     &   32 & 0 & 0 &  52 & 0 & 0       & 72 & 0 & 0               & 92 & 0 & 0 \\
 13 & 0 & 0     &   33 & 0 & 0 &  53 & 0 & 0       & 73 & 0 & 0               & 93 & 0 & 0 \\
 14 & 0 & 0      &  34 & 0 & 0 &  54 & 0 & 0       & 74 & 0 & 0               & 94 & 0 & 0 \\
 15 & 1 & 669.51  &  35 & 0 & 0&  55 & 0 & 0       & 75 & 0 & 0               & 95 & 0 & 0 \\
 16 & 0 & 0       &  36 & 0 & 0&  56 & 0 & 0       & 76 & 0 & 0               & 96 & 1 & 1105.77 \\
 17 & 1 & 806.61  &  37 & 0 & 0&  57 & 0 & 0       & 77 & 0 & 0               & 97 & 0 & 0 \\
 18 & 1 & 401.80  &  38 & 0 & 0&  58 & 0 & 0       & 78 & 0 & 0               & 98 & 0 & 0 \\
 19 & 0 & 0       &  39 & 0 & 0&  59 & 0 & 0       & 79 & 0 & 0               & 99 & 1 & 200 \\
 20 & 0 & 0       &  40 & 0 & 0&  60 & 0 & 0       & 80 & 0 & 0               & 100 & 0 & 0\\
 \hline\hline
\end{tabular}
\end{center}
\end{table}

Figure \ref{fig2} shows the complete number of claims and the
total claim amount concerning these claims. It can be seen that
the largest claim values appear in the case of single claims, while
these values fall with larger numbers of claims.
\begin{figure}[h]
\begin{center}
\includegraphics*[scale=1]{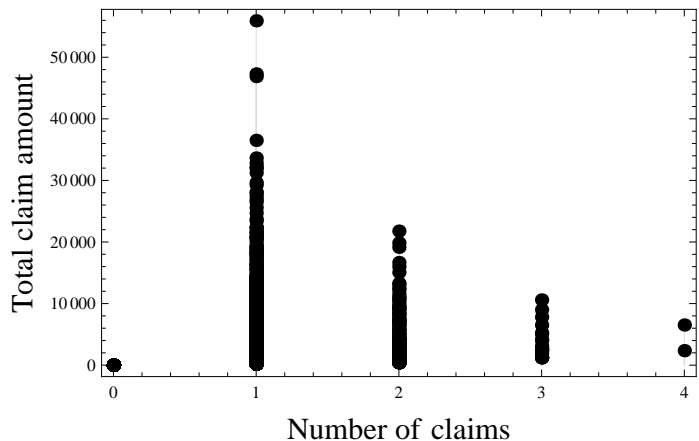}
\caption{Number of claims and their sizes\label{fig2}}
\end{center}
\end{figure}

\begin{table}[htb]
\caption{Some descriptive data of claims and claim size for the
data set\label{tab2}}
\begin{center}
\begin{tabular}{rrr}\hline\hline\\
& {\begin{tabular}{c}Number of\\claims\end{tabular}} & {\begin{tabular}{c}Total claim\\amount\end{tabular}}\\
\cline{2-3}\\
Mean & 0.072 & 137.27\\
Standard deviation & 0.278 & 1056.30\\
$\min$ & 0 & 0\\
$\max$ & 4 & 55922.10 \\ \\
\hline\hline
\end{tabular}
\end{center}
\end{table}

In this regard, the compound Poisson model has been traditionally
considered when the size of a single claim is modeled by an
exponential distribution, chiefly because of the complexity of the
collective risk model under other probability distributions such
as Pareto and log-normal distributions.

Perhaps the most well-known aggregate claims model is the obtained
when the primary and secondary distribution are the Poisson and the
exponential distributions, respectively.  In this case (see Rolski
et al. (1999), among others)  the distribution of the random
variable total claim amount is given by
\begin{eqnarray*}
f_S(x;\alpha,\lambda) =\sqrt{\frac{\lambda\alpha}{x}}\exp(-\lambda-\alpha x)
I_{1}\left(2\sqrt{\lambda\alpha
x}\right),\,x>0,\label{verosPoisson}
\end{eqnarray*}
while $f_S(0;\alpha,\lambda)=\exp(-\lambda)$. Here, $\lambda>0$ and $\alpha>0$
are the parameters of the Poisson and exponential distributions,
respectively and
\begin{eqnarray}
I_{\nu}(z)=\sum_{k=0}^{\infty}\frac{(z/2)^{2k+\nu}}{\Gamma(k+1)\Gamma(\nu+k+1)},\,
z\in \mathbb{R},\,\nu\in \mathbb{R},\label{veroNB}
\end{eqnarray}
represents the modified Bessel function of the first kind.

Additionally, the negative binomial distribution with parameters
$r> 0$ and $0 < p <$ 1 could also be assumed as primary
distribution and the exponential distribution as the secondary
distribution. In this case, the PDF of the random variable total
claim amount (see Rolski et al. (1999)) is now given by the
expression
\begin{equation*}\label{verosNB}
f_s (x;\alpha,r,p) =\alpha r p^r(1-p)\exp(-\alpha x){}_{1}F_{1}[1+r;2;\alpha(1-p)x],\, x>0,
\end{equation*}
where ${}_{1}F_{1}[\cdot;\cdot;\cdot]$ is the confluent
hypergeometric function and $f_S(0;\alpha,r,p)=p^{r}$.

When $r=1$ in (\ref{verosNB}) we get the PDF of the total claim
amount when the geometric distribution is considered as primary
distribution and the exponential distribution as secondary. This
results in
\begin{eqnarray*}
f_S(x;\alpha,p)=\alpha p(1-p)\exp(-\alpha p x),\quad x>0,
\end{eqnarray*}
with $f_S(0;\alpha,p)=p$.

Moment estimators can be obtained by equating the sample moments
to the population moments. Furthermore, the parameters of the
different models of the total claim amount can be estimated via
maximum likelihood. We only present the case of
Pareto-geometric case. To do so, consider a random sample
$\{x_1,x_2,\dots,x_n\}$. The log--likelihood function can be
written as
\begin{eqnarray*}
\ell(\alpha,\beta,p) &=& n_0\log p+(n-n_0)\left[
\log\alpha+\log p\right.\\
&&\left.+\log(1-p)+\alpha\log\beta\right]
-(\alpha+1)\sum_{x_i>0} \log(\beta+p x_i),\nonumber\\
\label{lik}
\end{eqnarray*}
where $n_0$ is the number of zero-observations and $n-n_0$ is the
number of non--zero sample observations, where $n$ is the sample
size.

The equations from which we get the maximum likelihood estimates
cannot be solved explicitly. They must be solved either by
numerical methods or by directly maximizing the log--likelihood
function. This was the method carried out here. Since the global
maximum of the log--likelihood surface is not guaranteed,
different initial values in the parameter space were considered
as a seed point. We use the {\tt FindMaximum} function of
Mathematica software package v.10.0 (Wolfram (2003)). Different
maximization algorithms such as {\tt Newton}, {\tt PrincipalAxis}
and {\tt QuasiNewton} were used to ensure that the same estimates
are obtained. Finally, the variances of the maximum likelihood
estimates can be estimated by the  diagonal elements of the
inverse matrix of negative second derivatives of the
log-likelihood function, evaluated at the maximum likelihood
estimates. When hypergeometric functions appear, these were
replaced by their series representation by taking one hundred terms
in the sum. This facilitates the computation of the Hessian
matrix.

A summary of the results obtained is shown in Table \ref{tab3}. In
this Table the estimated parameter values are presented
together with their standard errors in parenthesis, the AIC and
the CAIC. Bozdogan (1987)  proposed a corrected version of AIC in
an attempt to overcome the tendency of the AIC to overestimate the
complexity of the underlying model. Bozdogan (1987) observed that
Akaike Information Criteria (AIC) (see Akaike (1973)) does not
directly depend on sample size and as a result lacks certain
properties of asymptotic consistency. In formulating CAIC, a
correction factor based on the sample size is employed to
compensate for the overestimating nature of AIC. The CAIC is
defined as $\mbox{CAIC}=-2\log\ell+(1+\log n)\,k$, where again
$\ell$ and $k$ refers to the likelihood under the fitted model and
the number of parameters, respectively and $n$ is the sample size.
As we can see, AIC differs from CAIC in the second term which now
takes into account the sample size $n$. Again, models that minimize
the Consistent Akaike Information Criteria are selected. Our results
point out that the compound Pareto model outperforms the Poisson-exponential
and the geometric-exponential models, which have been widely considered in
the actuarial literature when parametric models are used.

\begin{landscape}
\begin{table}[htbp]
\caption{Summary of results for the different models considered\label{tab3}}
\begin{center}
\begin{tabular}{llccccccc}\hline\hline
\multicolumn{2}{l}{Distribution} &  && & & \\    \cline{1-2}
Primary & Secondary & $\widehat r$ & $\widehat p$ & $\widehat \lambda$ & $\widehat \alpha$ & $\widehat \beta$ & AIC & CAIC\\   \hline
Poisson & Exponential &  & & 0.12057 & 0.87832 && 51402.60 & 51422.80\\
 & &  & & $(0.00104)$ & $(0.00759)$ & & \\
Geometric & Exponential  &  & 0.93186 & 0.53273 &&& 49495.40 & 49515.60\\
&& & $(0.00097)$ & $(0.00785)$ && & \\
Negative Binomial & Exponential & 0.51168 & 0.87090 && 0.55250 && 49487.20 & 49517.60\\
 && $(0.00000)$ & $(0.0000)$ && $(0.00471)$ &&&\\
Poisson & Pareto  && & 0.07058 & 2.04828 & 2.13071 & 48229.50 & 48259.90\\
 &&&& $(0.00102)$ &  $(0.00974)$ & $(0.04879)$ && \\
 Geometric & Pareto&  & 0.93186 && 2.04655 & 2.05481 & 48229.60 & 48260.00\\
 &&& $(0.00097)$ && $(0.08828)$ & $(0.12407)$ &\\
 Negative Binomial & Pareto & 0.31749 & 0.80067 && 2.05542 & 1.91539 & 48232.10 & 48272.60\\
 && $(0.21505)$ & $(0.12052)$&& $(0.09000)$ & $(0.17324)$ &&\\
  \hline\hline
\end{tabular}
\end{center}
\end{table}
\end{landscape}

Figure \ref{fig2} shows the PDF of the four models considered
 using the parameter estimated given in Table \ref{tab3}.
As we can see,the new compound models have a larger
right tail than the traditional models based on the use of the
exponential distribution.
\begin{figure}[h]
\begin{center}
\includegraphics*[scale=.5]{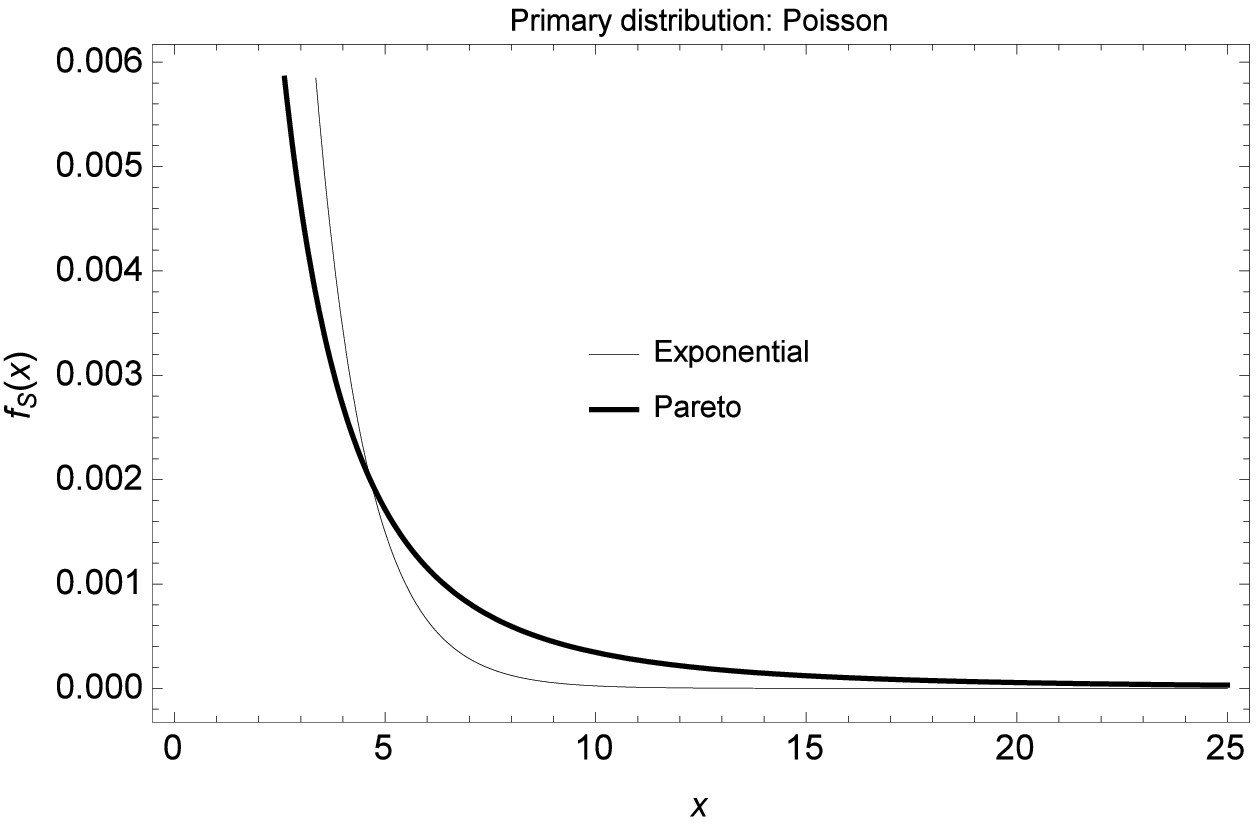}\hspace{0.15cm}
\includegraphics*[scale=.5]{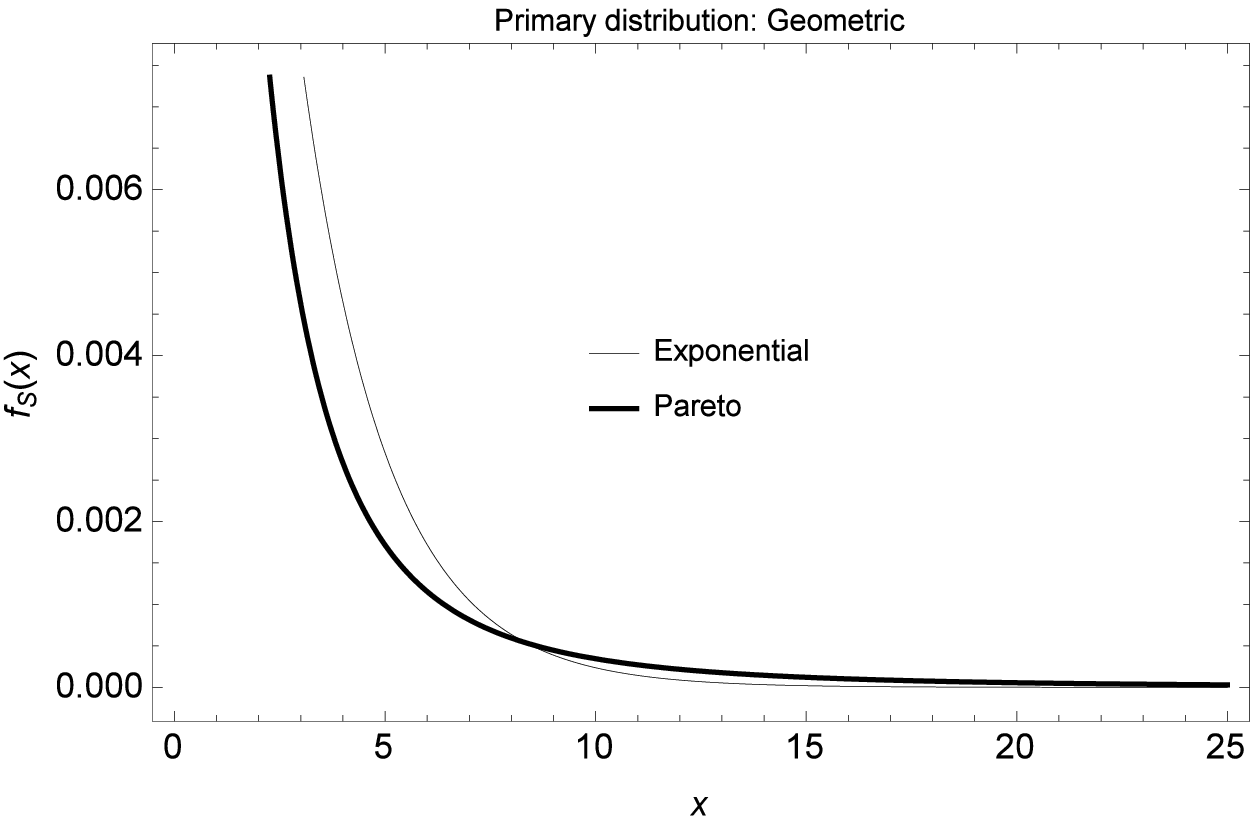}\\
\includegraphics*[scale=.5]{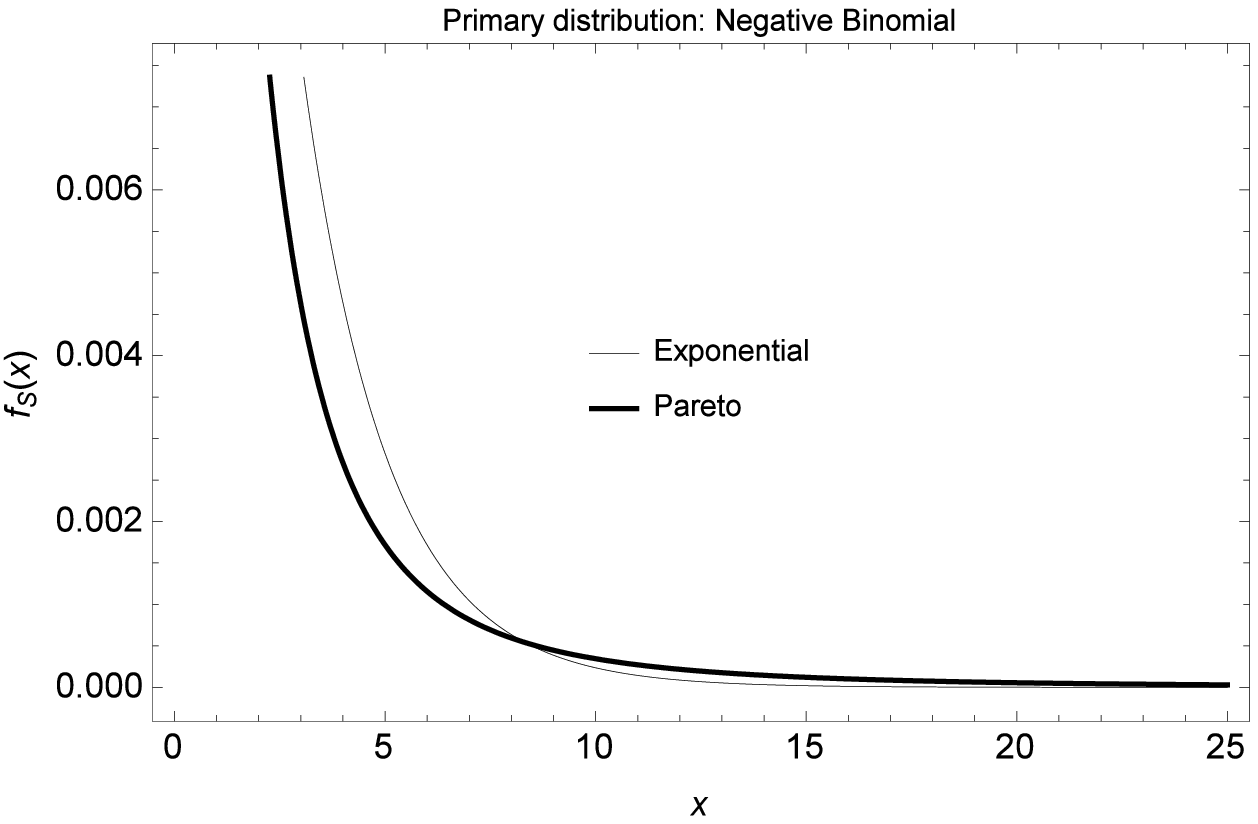}
\caption{PDF of the compound distributions for the estimated parameters of the models} \label{fig2}\end{center}
\end{figure}

Parameter estimates presented in Table \ref{tab3} have been used to
calculate the right-tail cumulative probabilities for different
values of $x$ as displayed in Table \ref{tab4}. As it can be
inferred from this Table, for values of $x<5$ the compound
exponential--Poisson and exponential--geometric models has a
slightly better performance than the new compound models in terms
of the decreasing probabilities they generate. Nevertheless, the
opposite result is obtained for $x>5$. Furthermore, new compound models tend
to zero slower than the models based on the exponential secondary
distribution.

\begin{landscape}
\begin{table}[htbp]
\caption{Right-tail cumulative probabilities of the aggregate
claim distribution for the estimated parameters of the
models\label{tab4}}
\begin{center}
\begin{tabular}{ccccccc}\hline\hline \\
& \multicolumn{2}{c}{Primary distribution: Poisson} &
\multicolumn{2}{c}{Primary distribution: Geometric} &\multicolumn{2}{c}{Primary distribution: Neg. Bin.}\\
& \multicolumn{2}{c}{\underline{Secondary distribution}} & \multicolumn{2}{c}{
\underline{Secondary distribution}} &\multicolumn{2}{c}{\underline{Secondary distribution}}\\
$x$ & Exponential & Pareto & Exponential & Pareto & Exponential & Pareto\\ \hline \\
 1 &  0.0496829  & 0.0317014 & 0.0414796 & 0.0316985  &     0.0415048  &   0.0317054   \\
 2 &  0.0217140  & 0.0181808 & 0.0252488 & 0.0181835   &    0.0252360 &    0.0181934   \\
 3 &  0.0094826  & 0.0117457 & 0.0153690 & 0.0117506  &     0.0153486  &   0.0117580   \\
 4 &  0.0041379  & 0.0081956 & 0.0093551 & 0.0082009   &    0.0093377 &    0.0082053     \\
 5 &  0.0018043  & 0.0060350 & 0.0056945 & 0.0060403  &     0.0056824 &    0.0060423     \\
 6 &  0.0007862  & 0.0046246 & 0.0034662 & 0.0046296  &     0.0034589 &    0.0046298     \\
 7 &  0.0003423  & 0.0036541 & 0.0021099 & 0.0036587 &      0.0021060 &    0.0036577     \\
 8 &  0.0001490  & 0.0029583 & 0.0012843 & 0.0029625 &      0.0012826 &    0.0029607     \\
 9 &  0.0000648  & 0.0024428 & 0.0007817 & 0.0024467 &      0.0007814 &    0.0024443      \\
 10 & 0.0000281  & 0.0020504 & 0.0004758 & 0.0020540 &      0.0004761 &    0.0020513      \\
 11 & 0.0000122  & 0.0017450 & 0.0002896 & 0.0017482 &      0.0002902 &    0.0017453      \\
 12 & 5.3132E--6 & 0.0015026 & 0.0001763 & 0.0015056 &      0.0001769 &    0.0015026      \\
 13 & 2.3057E--6 & 0.0013072 & 0.0001073 & 0.0013099 &      0.0001078 &    0.0013069      \\
 14 & 1.0000E--6 & 0.0011473 & 0.0000653 & 0.0011498 &      0.0000658 &    0.0011468      \\
 15 & 4.3357E--7 & 0.0010149 & 0.0000397 & 0.0010172 &      0.0000401 &    0.0010142      \\
 16 & 1.8787E--7 & 0.0009040 & 0.0000242 & 0.0009061 &      0.0000244 &    0.0009031      \\
 17 & 8.1375E--8 & 0.0008102 & 0.0000147 & 0.0008122  &     0.0000149 &    0.0008093      \\
 18 & 3.5230E--8 & 0.0007302 & 8.9686E--6 & 0.0007321  &    9.1270E--6 &   0.0007292      \\
 19 & 1.5246E--8 & 0.0006614 & 5.4592E--6 & 0.0006631 &     5.5729E--6 &   0.0006603     \\
 20 & 6.5952E--9 & 0.0006018 & 3.3230E--6 & 0.0006035 &     3.4034E--6 &   0.0006007      \\
\hline\hline
\end{tabular}
\end{center}
\end{table}
\end{landscape}

\section{Conclusions}\label{section7}

We have obtained closed expressions for the probabilistic distribution and several risk measures, modeling aggregated risks with multivariate dependent Pareto distributions.

We have worked with the dependent multivariate Pareto type II. For the case of the individual risk model, we have obtained the PDF of the aggregated risks, which corresponds to a second kind beta distribution. Then, we have considered the collective risk model based on dependence. We have studied some relevant collective models with Poisson, negative binomial and logarithmic distributions as primary distributions. For the collective Pareto-Poisson model, the PDF is a function of the Kummer confluent hypergeometric function, and in the Pareto-negative binomial is a function of the Gauss hypergeometric function.

Finally, using the data set based on one-year vehicle insurance policies taken out in 2004-2005 (Jong and Heller, 2008), we have concluded that our collective dependent models outperform the classical collective models defined by the Poisson-exponential and the geometric-exponential distributions in terms of the AIC and CAIC statistics.

\section*{Acknowledgements}

The authors thanks to the Ministerio de Econom\'ia y Competitividad (projects ECO2013-48326-C2-2-P, JMS, FP, VJ and ECO2013-47092 EGD) for partial support of this work.


\section*{Appendix}

\noindent{\bf Proof of \ref{proposition1}}
The proof is based on the fact that the random variables $(X_1,\dots,X_n)^\top$ are increasing functions of independent random variables and as a consequence they are associated random variables (Esary et al. 1967).\\

\noindent{\bf Proof of Theorem \ref{theorem1}}
We have,
$$S_n=X_1+\dots+X_n=\beta\frac{Y_1+\dots+Y_n}{Y_\alpha},$$
and the distribution of the numerator is a ${\cal G}a(n)$ and the denominator is ${\cal G}a(\alpha)$, where both random variables are independent. Consequently, the ratio is a second kind beta distribution with PDF given by (\ref{PDFindividual}).\\

\noindent{\bf Proof of Theorem \ref{theorem2}}
The proof is direct,
\begin{eqnarray*}
F_{S_N}(x)&=&\Pr(S_N\le x)\\
&=&\sum_{n=0}^\infty\Pr(S_N\le x|N=n)\Pr(N=n)\\
&=&\sum_{n=0}^\infty p_nF_X^{(n)}(x),
\end{eqnarray*}
where now $F_X^{(n)}(x)$ represents the CDF of the convolution of the $n$ dependent claims $(X_1,\dots,X_n)$.\\

\noindent{\bf Proof of Lemma \ref{lemma3}}
The tail moments can be expressed as,
$$E[X^r|X>a]=\frac{\int_a^\infty x^rdF_X(x)}{1-F_X(a)},$$
where $a=\mbox{VaR}[X;u]$. Using the incomplete distribution of the second kind beta distribution we obtain the result.\\

\noindent{\bf Proof of Lemma \ref{lemma2}}
The formula for $E(S_N)$ is direct. Formula for $var(S_N)$ can be obtain using the identity $var(S_N)=E[var[S_N|N]]+var[E[S_N|N]]$.\\

\noindent{\bf Proof of Theorem \ref{theorem3}}
If $x=0$, $f_{S_N}(0)=\Pr(N=0)=e^{-\lambda}$. If $x>0$ and calling $z=\frac{\lambda x/\beta}{1+x/\beta}$,
\begin{eqnarray*}
f_{S_N}(x)&=&\sum_{n=1}^\infty f_X^{(n)}(x)\Pr(N=n)\\
&=&\sum_{n=1}^\infty\frac{x^{n-1}}{\beta^nB(n,\alpha)(1+x/\beta)^{n+\alpha}}\frac{e^{-\lambda}\lambda^n}{n!}\\
&=&\frac{x^{-1}e^{-\lambda}}{\Gamma(\alpha)(1+x/\beta)^\alpha}\sum_{n=1}^\infty\frac{\Gamma(n+\alpha)}{\Gamma(n)n!}\left(\frac{\lambda x/\beta}{1+x/\beta}\right)^n\\
&=&\frac{x^{-1}e^{-\lambda}}{\Gamma(\alpha)(1+x/\beta)^\alpha}z\Gamma(\alpha+1){}_1F_1[1+\alpha;2;z],
\end{eqnarray*}
and we obtain (\ref{PDFParetoPoisson}).\\

\noindent{\bf Proof of Corollary \ref{corolario11}}
The proof is direct using Theorem \ref{theorempbn} and taking into account that,
$${}_2F_1\left[2,1+\alpha;2;z\right]=(1-z)^{-\alpha-1}.$$

\section*{References}

\begin{description}

\item Akaike, H. (1973). Information Theory and an Extension of the Maximum Likelihood Principle. In: B.N. Petrov and
    F. Csaki (eds.) 2nd International Symposium on Information Theory, 267-281.

\item Albrecher, H., Teugels, J. (2006). Exponential behavior in the presence of dependence in risk theory. \textit{Journal of Applied Probability}, 43, 265-285.

\item Arnold, B.C. (1983). \textit{Pareto Distributions}. International Cooperative Publishing House, Fairland, MD.

\item Arnold, B.C. (1987). Bivariate distributions with Pareto conditionals. \textit{Statistics and Probability Letters}, \textbf{5}, 263-266.

\item Arnold, B.C. (2015). \textit{Pareto Distributions}, Second Edition. Chapman \& Hall/CRC Monographs on Statistics \& Applied Probability, Boca Rat\'on, FL.

\item Arnold, B.C., Castillo, E., Sarabia, J.M. (1993). Multivariate distributions with generalized Pareto conditionals. \textit{Statistics and Probability Letters}, \textbf{17}, 361-368.

\item Arnold, B.C., Castillo, E., Sarabia, J.M. (2001). Conditionally Specified Distributions: An Introduction (with discussion). \textit{Statistical Science}, 16, 151-169.

\item Asimit, A.V., Furman, E., Vernic, R. (2010). On a multivariate Pareto distribution. \textit{Insurance: Mathematics and Economics}, \textbf{46}, 308-316.

\item Boudreault, M., Cossette, H., Landriault, D., Marceau, E. (2006). On a risk model with dependence between interclaim arrivals and claim sizes. \textit{Scandinavian
Actuarial Journal}, 2006, 301-323.

\item Bozdogan, H. (1987). Model selection and Akaike's Information Criterion (AIC): The general theory and its analytical extensions. {\it Psychometrika}, 52,
    345-370.

\item Chiragiev, A., Landsman, Z. (2009). Multivariate flexible Pareto model: Dependency structure, properties and characterizations. \textit{Statistics and Probability Letters}, \textbf{79}, 1733-1743.

\item Cossette, H., Cot\'e, M.-P., Marceau, E., Moutanabbir, K. (2013). Multivariate distribution defined with Farlie-Gumbel-Morgenstern copula and mixed Erlang marginals: Aggregation and capital allocation. \textit{Insurance: Mathematics and Economics}, 52, 560-572.

\item Cossette, H., Landriault, D., Marceau, E. (2004). Compound binomial risk model in a markovian environment. \textit{Insurance: Mathematics and Economics}, 35, 425-443.

\item Cossette, H., Marceau, E., Marri, F. (2008). On the compound Poisson risk model with dependence based on a generalized Farlie–Gumbel–Morgenstern copula. \textit{Insurance: Mathematics and Economics}, 43, 444-455.

\item Esary, J.D., Proschan, F., Walkup, D. W. (1967). Association of random variables, with applications. \textit{Annals of Mathematical Statistics}, \textbf{38}, 1466-1474.

\item Genest, G., Marceau, E., Mesfioui, M. (2003). Compound Poisson approximations for individual models with dependent risks. \textit{Insurance: Mathematics and Economics}, 32, 73-91.

\item Gerber, H.U. (1988). Mathematical fun with the compound binomial process. \textit{ASTIN Bulletin}, 18, 161-168.

\item G\'omez-D\'eniz, E., Calder\'in, E. (2014). Unconditional distributions obtained from conditional specifications models with applications in risk theory. \textit{Scandinavian Actuarial Journal}, 2014, 7, 602-619.

\item Guill\'en, M., Sarabia, J.M., Prieto, F. (2013). Simple risk measure calculations for sums of positive random variables. \textit{Insurance: Mathematics and Economics}, 53, 273-280.

\item Jong, P. and Heller, G. (2008). {\it Generalized Linear Models for Insurance Data}. Cambridge University Press.

\item Kaas, R., Goovaerts, M., Dhaene, J., Denuit, M. (2001). \textit{Modern Actuarial Risk Theory}. Kluwer Academic Publishers, Boston.

\item Klugman, S.A., Panjer, H.H., Willmot, G.E. (2008). \textit{Loss Models. From Data to Decisions}, third edn. John Wiley, New York.

\item Ramsay, C.M. (2006). The distribution of sums of certain i.i.d. Pareto variates. \textit{Communications in Statistics, Theory and Methods}, 35, 395-405.

\item Ramsay, C.M. (2009). The distribution of compound sums of Pareto distributed losses. \textit{Scandinavian Actuarial Journal}, 2009, 27-37.

\item Rolski, T., Schmidli, H. Schmidt, V. and Teugel, J. (1999). {\it Stochastic Processes for Insurance and Finance.} John Wiley and Sons, New York.

\item Sarabia, J.M., Guill\'en, M. (2008). Joint modelling of the total amount and the number of claims by conditionals. \textit{Insurance: Mathematics and Economics}, 43 466-473.

\item Wolfram, S. (2003). {\it The Mathematica Book}. Wolfram Media, Inc.

\end{description}

\end{document}